\documentclass{llncs}

\usepackage[pdftex]{graphicx}
\usepackage{epstopdf}        
     
\usepackage{makeidx}  
\usepackage{amsmath}
\usepackage{amsfonts,amssymb}
\usepackage{clrscode3e}        
\usepackage{multirow}

\begin{document}
\frontmatter          
\pagestyle{headings}  
\addtocmark{Graph traversals} 
\title{Graph traversals\\ associated with iterative methods\\ for solving systems of linear equations}
\titlerunning{Graph traversals associated with iterative methods for solving SLAE}  
%
\author{A.V. Prolubnikov}
\authorrunning{A.V. Prolubnikov} 
%
\tocauthor{A.V. Prolubnikov}
\institute{Novosibirsk State University, Novosibirsk, Russian Federation\\
\email{a.v.prolubnikov@mail.ru}
}
\maketitle              

\begin{abstract}

To solve many problems on graphs, graph traversals are used, the usual variants of which are the depth-first search and the breadth-first search. Implementing a graph traversal we consequently reach all vertices of the graph that belong to a connected component. The~breadth-first search is the usual choice when constructing efficient algorithms for finding connected components of~a~graph. Methods of simple iteration for solving systems of linear equations with modified graph adjacency matrices can be considered as graph traversal algorithms if we use properly specified right-hand side. These traversal algorithms, generally speaking, turn out to be non-equivalent neither to the depth-first search nor the breadth-first search. The example of such a~traversal algorithm is the~one associated with the Gauss-Seidel method. For an arbitrary connected graph, to visit all its vertices, the algorithm requires not more  iterations than that is required for~breadth-first search. For a large number of instances of the problem, fewer iterations will be required.

\end{abstract}

\section*{Introduction}

Many problems from applications related to infrastructure reliability (transport networks, data networks, large integrated circuits, etc.) formulated as~graph connectivity problems. These are problems such as finding connected components of~a~graph, finding its articulation points, bridges, etc.

Let $G$ be a simple labeled graph, that is, an undirected unwheighted graph without multiple edges and loops, $V$ is the set of its vertices, $|V|\!=\!n$, $E$ is the set of its edges, $|E|\!=\!m$. We will assume that the vertices of the graph are labeled (numbered) in some, generally speaking, arbitrary order from $1$ to $n$. A {\it connected component} of an undirected graph is a connected subgraph that is not part of any larger connected subgraph. The~connected component is defined by the set of its vertices.

{\it Graph traversal} is the iterative process during which, begining from some {\it starting vertex}, transitions along the edges of~the~graph are implemented. The traversal completed if all vertices of the connected component have been visited. By~implementing graph traversal by an algorithm we mean the iterative process of obtaining the vertices through which the~sequential transitions along graph edges are carried out.

Graph traversals are natural tool that used to solve connectivity problems on graphs. Common options for graph traversal algorithms are the depth-first search and the breadth-first search. 

The~{\it depth-first search (DFS)} strategy \cite{Cormen} is to go as deep into a graph as possible. The search algorithm is described recursively: we go through all the edges incident to the current vertex; if an edge leads to a vertex that has not been visited previously, then we run the algorithm from this unconsidered vertex, and then return and continue to iterate over the edges. Return occurs if~there are no edges left which is incident to the current vertex and that lead to yet unvisited vertices of the connected component.

Usually, the {\it breadth-first search (BFS)} \cite{Cormen} used to find connected components computationally efficiently for large graphs. Implementing BFS, begining from a starting vertex that belongs to some connected component, we perform transitions to yet unvisited vertices that adjacent to the vertices visited at the previous iteration. At the process of such iterations, we construct {\it reachability tree} for the connected component of a graph. At the $(k+1)$-th level of this tree, there are vertices which are reached from the starting vertex after sequential transitions along $k$ edges of some simple chain in the graph. The construction of the reachability tree for a connected component can be done in $\ell_{\max}$ iterations, where $\ell_{\max}$ is the length of the shortest simple chain which connects starting vertex to the vertex farthest from~it.

BFS was first used by K.~Zuse in 1945, but was not published until 1972 \cite{Zuse}. Subsequently, this algorithm was rediscovered in 1959 by E.~Moore as an algorithm for finding the shortest path in a labyrinth \cite{Moore} and was later independently rediscovered by C. Lee in the context of the problem of routing conductors on printed circuit boards~\cite{Lee}. 

Implementations and generalizations of BFS for the case of weighted graphs underlie many algorithms for solving discrete analysis and discrete optimization problems. For example, we do such kind of a graph traversal to find a tree of shortest paths to vertices of a graph, to check a graph for bipartiteness, to~compute the maximum flow in~a~flow network and for many other problems.

\smallskip

For a graph $G$, let $s\!\in\!V$ be the starting vertex. {\it Algebraic BFS} is an implementation of BFS through sequential multiplication of the vector $x^{(k)}$ by the graph adjacency matrix $A$: $$x^{(k+1)}=Ax^{(k)},\eqno(1)$$ where $x^{(0)}\!=\!e_s$, $e_s$ is the $s$-th unit vector of the standard basis in $\mathbb{R}^n$, i.e., all of its components are zero except the $s$-th component. At each iteration, some components of the multiplied vector become non-zero. The indices of such components of $x^{(k)}$ are numbers of the vertices visited after the $k$-th iteration of BFS has been implemented.

Algebraic BFS is implemented in software libraries designed to most effectively leverage the features of modern computer architectures for parallel computing and memory optimization. They are focused on working with large sparse graphs which are most often found in applications. Such low-level implementations in practice can be faster than implementations of theoretically optimal combinatorial BFS \cite{BeamerAsanovicPatterson,BuckerSohr,BulucMattsonMcMillanMoreiraYang,AzadBuluc,YangBulucOwens,YangWangOwens}. Since computational complexity of BFS is of $O(m+n)$, for sparse graphs, that is those for which $m\!=\!O(n)$, it allows to obtain algorithms with linear computational complexity \cite{Burkhardt} which is the best possible for the problem. 

And although computations (1) can be efficiently parallelized for computing of the current level of reachability tree, due to its inevitable sequential nature, there are no sublinear implementations of BFS, i.e., there is no implementations with computational complexity of $O(n^{c})$, where $0\!<\!c\!<\!1$.

\smallskip

So, there are two fundamentally different computational approaches to numerical realization of graph traversal algorithms. These are com\-bi\-na\-torial and algebraic (linear-algebraic) approaches. 

When implementing a {\it combinatorial} approach, beginning from the starting vertex, the possible options for moving along the edges of a graph are considered consequently, and then we do transitions to yet unreached vertices along some selected edges.

Combinatorial traversal algorithms, being asymptotically optimal, in practice may be too slow for solving real applied problems with large graphs. The main part of the algorithms is a combinatorial search on possible options for transitions from the current vertex. After this search is performed, we do the transitions. But it turn out to be too slow in comparison to linear-algebaic transformation (1) that gives all possible transitions at once and can be implemented in~parallel.

For example, for the graph of a connectivity problem associated with real transport network of a large region, the presence of an edge means that the variable is included in some equation or inequality of a partially integer linear programming problem. The dimensions of the problem are $n\!=\!367\ 840$, $m\!=\!53\ 404\ 685$, and there are $224$ connected components in the graph. While it takes about $100$ min to find all the connected components using the linear transformation that performed by Gauss-Seidel iterations, it takes $48$ hours to solve the instance using basic combinatorial scheme of BFS from~\cite{wiki}.

When implementing an {\it algebraic} approach, at each iteration of the graph traversal we do some linear transformation of the {\it state vector} $x^{(k)}$ to perform an iteration of the traversal. Analysis of  components of the state vector allows us to determine the vertices visited after the iteration has been completed. A visit to a vertex $i\!\in\!V$ is registered if the value of the $i$-th state vector's component ceases to be zero after this linear transformation. An example of algebraic approach is algebraic BFS which uses the operation of multiplying the state vector by adjacency matrix (1) as the linear transformation. 

The graph traversal algorithms we consider use another linear transformations, different from multiplying a vector by adjacency matrix. Namely, this is a transformations implemented by a methods of simple iteration for solving a system of linear equations. Applied to the systems of linear algebraic equations (further abbreviated as {\it SLAE}) with modified graph adjacency matrices and properly specified right-hand side, variations of simple iteration methods can be considered as graph traversal algorithms. These numerical methods may provide two ways for implementing graph traversal, one of which is performed by the Jacobi method and it is equivalent to BFS, and the other performed by the Gauss-Seidel method and it gives a traversal algorithm which is not equivalent to~BFS.

Unlike DFS and BFS, applied to a graph traversal, Gauss-Seidel iterations leverage given vertex numbering in a graph. Numbering (labeling) of vertices is a mandatory parameter of a problem instance that cannot be ignored performing computations. We call a simple chain in graph as {\it correct chain} if it is a chain with ascending order of numbers of vertices in it. Leveraging these chains, the approach implemented by the Gauss-Seidel method gives a traversal algorithm which is neither DFS nor BFS. At each iteration, at first, one BFS iteration is performed, and then, if only these chains exist, transitions made along edges of the correct chains which are outgoing from the new reached at this iteration vertices.

For an arbitrary connected graph, to visit all its vertices, the~tra\-versal associated with the Gauss-Seidel method requires a number of iterations that is no greater than that is required for~BFS. For a large number of instances of the problem, fewer iterations will be required. If the graph has enough correct chains in it, even sequential implementation of this traversal algorithm can be faster than the parallel version of BFS.

\section{Iterative numerical methods for~solving SLAE\\ and~graph~traversals associated with them}

Let us consider iterations of the Jacobi method and iterations of the Gauss-Seidel method for solving a SLAE 
$$Ax=b.\eqno(2)$$ Both of these methods are variations of the simple iteration method. The iterations of the Jacobi method {\it (Jacobi iterations)} are as follows: $$x^{(k+1)}=b-D^{-1}Ax^{(k)},$$ where $D$ is a diagonal matrix with diagonal elements of $A$. That is, for the $i$-th component of the approximate solution vector after the $(k+1)$-th iteration, we have $$x_i^{(k+1)}=\frac{1}{a_{ii}}\biggl (b_i-\sum\limits_{j\neq i}a_{ij}x_j^{(k)}\biggr ).\eqno(3)$$ Iterations of the Gauss-Seidel method {\it (Gauss-Seidel iterations)} have the following form: $$(L+D){x}^{(k+1)}=-U {x}^{(k)}+b,$$ where $L$ and $U$ are matrices with, respectively, subdiagonal and supradiagonal elements of $A$, that is $$x_i^{(k+1)}=\frac{1}{a_{ii}}\biggl (b_i-\sum\limits_{j=1}^{i-1}a_{ij}x_j^{(k+1)}-\sum\limits_{j=i+1}^{n}a_{ij}x_j^{(k)}\biggr ).\eqno(4)$$ 

Using these methods to solve SLAE for graph matrices, in order to achieve solvability of the systems and for convergence of the methods, we modify adjacency matrix of a graph by replacing its zero diagonal elements with sufficiently large value $d\!>\!0$. If the matrix $A$ has diagonal dominance, that~is, the following condition is met: $$|a_{ii}|=d\ge\sum\limits_{i \neq j}|a_{ij}|$$ for all $i\!=\!\overline{1,n}$, and at least one of these inequalities is strict, then approximate solutions obtained at iterations (3) and (4) converge to the exact solution that exists and that is~unique for any right-hand side $b$. 

Consider SLAE (2), where $A$ is a modified graph adjacency matrix with diagonal dominance and $b\!=\!e_s$. The graph traversals algorithms that we consider further, are the ones that we obtain implementing the iterations (3) and (4). After performing not more than $n$ iterations (3) or (4) with $x^{(0)}\!=\!d\cdot e_s$ as initial vector, maybe without achieving convergence to the exact solution of the SLAE~(2), we obtain a sequence of approximate solutions $x^{(k)}$. Further, rather than approximate solutions of a SLAE, we shall consider $x^{(k)}$ as current value of the {\it state vector} $x^{(k)}$ for iteration $k$. 

Transitions between vertices of a graph can be registered using these computations. At the $(k+1)$-th iteration, the transition occurs to some vertex $i$ from one of the vertices reached at the $k$-th iteration if the following condition is met: $$(x_i^{(k)}=0)\ \mbox{and}\ (x_i^{(k+1)}\neq 0).\eqno(5)$$ That is, the vertex $i$ was not visited prior to the $(k+1)$-th iteration, and it is reached at this iteration. The {\it frontier} $\mathcal{F}^{(k+1)}$ for the $(k+1)$-th iteration is the set of the vertices for which corresponding components of the state vector cease to be zero at this iteration. $\mathcal{F}^{(0)}\!=\!\{s\}$.

Performing Jacobi iterations or Gauss-Seidel iterations with the given initial value of the state vector and the right-hand side of the SLAE (2), we sequentially obtain frontiers of traversals and, finally, visit all vertices of a connected graph.

All of the simple iteration methods for solving SLAE give the same two traversal algorithms in the sence of the sequentially obtained traversals. One of them is given by (3), and, as we shall see, it is equivalent to BFS, and the other is given by (4), and it is not equivalent to BFS.

\section{Using multiplication instead of division\\ in Jacobi and Gauss-Seidel iterations}

Since we do not need to achieve convergence to the exact solution of the SLAE (2) when performing iterations (3) and (4) to implement graph traversal, multiplication can be used instead of dividing in (3) and (4). Without affecting in any way the graph traversal algorithms considered below and the proofs carried out, it makes text more compact. After this trans\-for\-mation, the values $x_i^{(k+1)}$ become polynomials of $d$ but not of $1/d$.

Moreover, it makes practical sense to use multiplying instead of division in Jacobi and Gauss-Seidel iterations to perform graph traversal. Implementations of division with floating point machine numbers are several times slower than implementations of multiplication. So, after such a replacement, the time required for implementing graph traversal using the modified iterations will be shorter. 

Thus, further, considering adjacency graph matrices with initially zero diagonal elements replaced by some $d\!>\!0$, we consider iterations for the two methods as, respectively, (6) and (7):
$$x_i^{(k+1)}=\biggl (-b_i+\sum\limits_{j\neq i}a_{ij}x_j^{(k)}\biggr )(-d),\eqno(6)$$  
$$x_i^{(k+1)}=\biggl (-b_i+\sum\limits_{j=1}^{i-1}a_{ij}x_j^{(k+1)}+\sum\limits_{j=i+1}^{n}a_{ij}x_j^{(k)}\biggr )(-d).\eqno(7)$$ The value of $d$ can be arbitrary positive value.

\section{Combinatorial graph traversal algorithms}

A {\it chain in a graph} is a finite sequence of vertices in which each vertex except the last one is connected to the next one in the sequence by an edge. A chain may be considered as sequence of these edges. The {\it length of the chain} $c$ we denote as $\ell(c)$ which is the number of edges in $c$. A chain with the same initial and last vertex is called a {\it cycle}. We call a chain as a {\it correct} chain if its vertices' labels which are unique numbers, are given in ascending order, i.e., it is a chain $c\!=\{i_0,i_1,\ldots,i_k\}$, $i_j\!\in\!V$, such that $(i_{j-1}, i_j)\!\in\!E$, $i_{j-1}\!<\!i_j$, $j\!=\!\overline{1,k}$. Correct chains are {\it simple chains}, i.e., chains without self-intersections. A {\it walk} is a chain where vertices and edges both can be repeated in it. 

Let us call as {\it correct chain search} the traversal algorithm implemented by the iterations of the form~(7). For short, we will abbreviate it as {\it CCS}. We consider both BFS and CCS graph traversals as algorithms of the two types: combinatorial algorithms and algebraic algorithms. 

The similarity and the difference between BFS and CCS are as follows. As for BFS, performing iterations (7), at first, we perform transitions along the edges that is incident to vertices reached at the previous iteration. The difference between these graph traversal algorithms is~that if there are outgoing correct chains from the vertices visited at the current iteration, then at the same iteration, transitions through all vertices of~these correct chains will be done. 

This happens since, when performing the iterations (7), computations of individual components of the current state vector are performed using the values of the components which have been computed earlier in the course of the  current iteration. So, if vertex $j$ is adjacent to vertex $i$ and $j\!<\!i$, and the $j$-th component of the state vector ceases to be zero at the current iteration, then at the same iteration its value will be used for computing of the $i$-th component and it ceases to be a zero too if it was zero before. On the contrary, the BFS state vector is updated based only on the results of individual computations for each component of the state vector. It is good since it allows to perform these computations in parallel, but it not allows the frontier to be wider than neibourghood of the previous frontier vertices. For CCS, if there are correct chains that outgo from the neibourghood of the previous frontier, we traverse through them at the same iteration adding new vertices to the frontier. 

For $(k+1)$-th iteration of the combinatorial BFS, $\mathcal{F}^{(k+1)}$ is a set of vertices reached by traversing edges which are incident to vertices in $\mathcal{F}^{(k)}$ and which not traversed in the course of the previous $k$ iterations. Let $\mathcal{N}$ be this set. $\mathcal{C}^{(k)}$~is the set of vertices in $V$ visited after $k$ iterations of traversal algorithm. 

\begin{codebox}
\Procname{$\proc{\bf Combinatorial BFS}\ (G, s\!\in\!V): \mathcal{C};$}
\li $k\leftarrow 0$, $x^{(0)}\!\leftarrow\!e_s$, $\mathcal{F}^{(0)}\!\leftarrow\!\{s\}$, $\mathcal{C}^{(0)}\!\leftarrow\!\{s\}$;  
\li \While $\mathcal{C}^{(k)}\!\neq\!\mathcal{C}^{(k+1)}$:
\Do
\li $\mathcal{N}\leftarrow \bigl \{i\!\in V\!\setminus\!\mathcal{C}^{(k)}\,\bigl |\,(\exists j\in\!\mathcal{F}^{(k)})((i,j)\!\in E)\bigr \}$;
\li $\mathcal{F}^{(k+1)}\leftarrow \mathcal{N}$;
\li $\mathcal{C}^{(k+1)}\leftarrow\mathcal{C}^{(k)}\cup\mathcal{F}^{(k+1)}$;
\li $k\leftarrow k+1$;
\End				
\li $\mathcal{C}\leftarrow\mathcal{C}^{(k+1)}$;
\li \mbox{\bf result}$\ \leftarrow \mathcal{C}$. 		

\end{codebox}

Implementing iteration of CCS, in addition to traversing edges that incident to vertices in $\mathcal{F}^{(k)}$, we perform traversals of the correct chains that outgo from the vertices that reached through these incident edges. Let $C(\tilde{s},i)$ be the set of all correct chains outgoing from vertex $\tilde{s}\!\in\!\mathcal{N}$ and ending at vertex $i$.

\begin{codebox}
\Procname{$\proc{\bf Combinatorial CCS}\ (G, s\!\in\!V): \mathcal{C};$}
\li $k\leftarrow 0$, $x^{(0)}\!\leftarrow\!e_s$, $\mathcal{F}^{(0)}\!\leftarrow\!\{s\}$, $\mathcal{C}^{(0)}\!\leftarrow\!\{s\}$;  
\li \While $\mathcal{C}^{(k)}\!\neq\!\mathcal{C}^{(k+1)}$:
\Do
\li $\mathcal{N}\leftarrow \bigl \{i\!\in V\!\setminus\!\mathcal{C}^{(k)}\,\bigl |\, (\exists j\!\in\!\mathcal{F}^{(k)})((i,j)\in E)\bigr \}$;		    
				
\li $\mathcal{F}^{(k+1)}\leftarrow \mathcal{N}\cup \bigl\{i\!\in V\!\setminus\!(\mathcal{C}^{(k)}\cup\mathcal{N}) \,\bigl |\, (\exists \tilde{s}\!\in\!\mathcal{N})(C(\tilde{s},i)\neq\varnothing)\bigr \}$;
\li $\mathcal{C}^{(k+1)}\leftarrow\mathcal{C}^{(k)}\cup\mathcal{F}^{(k+1)}$; 
\li $k\leftarrow k+1$;
\End				
\li $\mathcal{C}\leftarrow\mathcal{C}^{(k+1)}$;
\li \mbox{\bf result}$\ \leftarrow \mathcal{C}$. 		
\end{codebox}

So, at the step 3 of both algorithms, we implement transitions along the edges incident to the vertices from the frontier $\mathcal{F}^{(k)}$ obtained at the previous iteration. Additional transitions through correct chains from $C(\tilde{s},i)$ for $\tilde{s}\!\in\!\mathcal{N}$ implemented at the step~4 of the $\proc{\bf Combinatorial CCS}$.

\section{Algebraic versions of the combinatorial BFS and CCS\\ implemented by Jacobi and Gauss-Seidel iterations} 

Let $\mbox{\bf F}$ denote the transformations of $x\!\in\!\mathbb{R}^n$ of the form (6) or (7). For a given starting vertex $s$, the following algorithm gives the connected component $\mathcal{C}$ to which $s$ belongs. 

\begin{codebox}
\Procname{$\proc{\bf Traversal of a connected component}\ (G, s\!\in\!V): \mathcal{C};$}
\li $k\leftarrow 0$, $x^{(0)}\!\leftarrow\!e_s$, $\mathcal{F}^{(0)}\!\leftarrow\!\{s\}$, $\mathcal{C}^{(0)}\!\leftarrow\!\{s\}$;  

\li \While $\mathcal{C}^{(k)}\!\neq\!\mathcal{C}^{(k+1)}$:
		   \Do
\smallskip
\li		    $x^{(k+1)}\leftarrow\mbox{\bf F}\bigl (x^{(k)}\bigr);$			
\smallskip
\li		    $\mathcal{F}^{(k+1)}\leftarrow\bigl\{i\!\in\!V(G)\, \bigl |\, (x_i^{(k)}=0) \wedge (x_i^{(k+1)}\neq 0)\bigl\}$;			
\li       $\mathcal{C}^{(k+1)}\leftarrow \mathcal{C}^{(k)}\cup\mathcal{F}^{(k+1)}$;
\li      $k\leftarrow k+1$;
			 \End					
       \End				
     \End				
\li $\mathcal{C}\leftarrow\mathcal{C}^{(k)}$;
\li \mbox{\bf result}$\ \leftarrow\mathcal{C}$. 		
\end{codebox}

For the case when $\mbox{\bf F}$ is the transformation of the form (6), the algorithm is the variation of algebraic BFS. For $\mbox{\bf F}$ of the form (7), it gives {\it algebraic CCS}. 

The algorithm that finds all connected components of a graph $G$ is given below. Connected components are denoted as $\mathcal{C}_i$, $i\!=\!\overline{1,K}$, where $K$ is the number of connected components of the graph $G$. 

\begin{codebox}
\Procname{$\proc{\bf Finding all connected components}\ (G):\{\mathcal{C}_1,\ldots,\mathcal{C}_K\};$}
\li $V'\leftarrow\varnothing$, $K\!\leftarrow\!1$;
\li \While $V'\!\neq\!V$:
\Do
\li \mbox{select} $s\!\in\!V\setminus V'$;
\li $\mathcal{C}_K\leftarrow \proc{\bf Traversal of a connected component}\ (G, s)$;
\li $V'\leftarrow V'\cup\mathcal{C}_K$;
\li $K\leftarrow K+1$;
\End
\li \mbox{\bf result}$\ \leftarrow \{\mathcal{C}_1,\ldots,\mathcal{C}_K\}$.
\end{codebox}

We find all connected components of a graph by sequentially choosing new starting vertices for traversals that we performs at the step 4 of the algorithm. 

\bigskip

Let us show that iterations (6) give us the graph traversal algorithm which is equivalent to the combinatorial BFS, and that iterations (7) give us the graph traversal algorithm which is equivalent to the combinatorial CCS. 

\section{Jacobi iterations as implementation\\ of the combinatorial BFS}

Considering BFS, for starting vertex $s$ and $i\!\in\!\mathcal{F}^{(k+1)}$, let $C(s,i)$ be the set of simple chains which connect $s$ with $i$. All of the chains, if there is any, have the length $k$, i.e., for all $c\!\in\!C(s,i)$ we have $\ell(c)\!=\!k$. 

\bigskip

\noindent{\bf Lemma 1.} {\it For iterations (6), we have $x_i^{(t)}\!=\!0$, $t\!=\!\overline{1,k}$, and} $$x_i^{(k+1)}\!=\!\sum\limits_{c\in C(s,i)}(-1)^{\ell(c)}d^{\ell(c)+1}\neq 0$$ {\it if and only if} $i\!\in\!\mathcal{F}^{(k+1)}$ {\it for the combinatorial BFS.} 

\bigskip

\noindent{\bf Proof.} We carry out the proof by induction on the number $k$ of iterations performed.

\smallskip

Let $k\!=\!1$. For $(s,i)\!\in\!E$, to obtain the values of $x_i^{(1)}$ from equations (6), the value of $x_s^{(0)}\!=\!d$ substituted into them, and we have $x_i^{(1)}\!=\!-d^2\!\neq\!0$ as a result. If the $i$-th equation in (6) does not contain $x_s^{(k)}$ in its right-hand side, that is $a_{is}\!=\!0$ and $(s,i)\!\not\in E$, then $x_i^{(1)}\!=\!0$. That is, Lemma~1 is true for this case.

Suppose Lemma~1 is true for the $k$-th iteration. That is, for $t\!<\!k$ and $i\!\in\!\mathcal{F}^{(k)}$, we have $x_i^{(t)}\!=\!0$, and $$x_i^{(k)}\!=\!\sum\limits_{c\in C(s,i)}(-1)^{\ell(c)}d^{\ell(c)+1}\neq 0.$$ Let us show that Lemma 1 is true for the $(k+1)$-th iteration too. 

Since Lemma~1 is true for the $k$-th iteration, for $i\!\neq\!s$, we have: $$x_i^{(k+1)}\!=\!\biggl (\sum\limits_{j=1}^n a_{ij}x_j^{(k)}\biggr )(-d)=\biggl (\sum\limits_{(i,j)\in E}\biggl (\sum\limits_{c\in C(s,j)}(-1)^{\ell(c)}d^{\ell(c)+1}\biggl ) \biggl )(-d)=$$ $$=\sum\limits_{(i,j)\in E}\biggl (\sum\limits_{c\in C(s,j)}(-1)^{\ell(c)+1}d^{\ell(c)+2}\biggl )=\sum\limits_{c\in C(s,i)}(-1)^{\ell(c)+1}d^{\ell(c)+2}\neq 0.$$

So, the condition (5) is met for $i\!\in\mathcal{F}^{(k+1)}$. Let us show that $x_i^{(k+1)}\!=\!0$ if $i\!\not\in\mathcal{F}^{(k+1)}$ and the vertex $i$ was not visted at previous $k$ iterations of the combinatorial BFS. This means that among the vertices $j$ adjacent to $i$ there are no ones for which there have been simple chains of length $k$ connecting $s$ and $j$ traversed in the course of the previous $k$ iterations. Since Lemma~1 is true for $k$, this means that for all such $j$ we have $x_j^{(k)}\!=\!0$. Therefore, $$x_i^{(k+1)}=\biggl (\sum\limits_{i=1}^na_{ij}x_j^{(k)}\biggr )(-d)=\biggl (\sum\limits_{(i,j)\in E}^nx_j^{(k)}\biggr )(-d)=0\cdot (-d)=0.$$ Lemma~1 is proven. $\qed$

\smallskip

Thus, by (5) and by Lemma~1, the iterations (6) produce the same frontiers $\mathcal{F}^{(k)}$, $k\!=\!1, 2, \ldots$, as iterations of the combinatorial BFS produce. This means that we obtain the same graph traversals during execution of the iterations. The Theorem~1 follows from that.

\bigskip

\noindent{\bf Theorem 1.} {\it For a given graph $G$ and starting vertex $s$, the combinatorial BFS and Jacobi iterations with $b\!=\!e_s$ and $x^{(0)}\!=\!d\cdot e_s$ give the same traversal.}

\section{Gauss-Seidel iterations as implementation\\ of the combinatorial CCS}

\subsection{Computing components of the state vector using traversals\\ of the chains outgoing from the current frontier's verices}

Considering the $(k+1)$-th iteration of the combinatorial CCS and iteration of the form (7) with the same number, we will use the following notations. Unlike to the previous section, we denote as $C(\tilde{s},i)$ the set of simple chains connecting the vertices $\tilde{s}\!\in\!\mathcal{F}^{(k)}$ with vertices $i\!\in\!\mathcal{F}^{(k+1)}$. Along edges of these chains transitions are made at $(k+1)$-th iteration of the combinatorial CCS. Let $c+(j,i)$ denotes the chain obtained from $c$ by connecting its last vertex $j$ to vertex $i$ by the edge $(j,i)\!\in\!E$.

All computations in iterations (6) and (7) can be decomposed into computations that implemented in accordance with traversals of individual chains (not necessary simple) connecting the starting vertex and the vertex $i$, for which the value $x_i^{(k+1)}$ is computing. We shall consider only simple chains since it will be enough for the proofs of the followed lemma and theorem.  Indeed, when implemeting computations of the form (6) or (7), the computations for some set of walks from starting vertex to other vertices are constitute the set of all individual arithmetic operations implemented in the course of one iteration.

Chains that can be traversed in the course of one iteration of combinatorial CCS, that is, chains $c\!=\!\{i_0,i_1,\ldots,i_{\ell(c)}\}\!\in\!C (\tilde{s},i)$, where $i_0\!=\!\tilde{s}$, are chains of the two types. A chain $c\!\in\!C(\tilde{s},i)$ is a {\it chain of type~(I)} if $i_0\!<\!i_1$, and {\it the chain type is (II)} if $i_0\!>\!i_1$. For both type chains, maybe except the first edge $(i_0,\!i_1)$, we have $i_{j-1}\!<\!i_j$ for $j\!=\!\overline{1,\ell(c)}$. Transition along the edge $(i_0,i_1)$ is implemented at the step 3 of the combinatorial CCS and transitions along subsequent edges of the chain are performed at the step 4.

For the vertices $\tilde{s},i\!\in\!V$, let $\ell(\tilde{s},i)\!=\!\max\{\ell(c)\, |\, c\in C(\tilde{s},i)\}$. $\ell(\tilde{s},i)$ is the maximum length of a simple chain outgoing from $\tilde{s}\!\in\!\mathcal{F}^{(k)}$ along edges of which the vertex $i\!\in\!\mathcal{F}^{(k+1)}$ is reached at the $(k+1)$-th iteration.

Let $i_0\!=\!\tilde{s}$, $i_{\ell(c)}\!=\!i$ in the chain $c\!=\!\{i_0,\ldots,i_{\ell(c)}\}$ and $v_{\tilde{s}}$ is the value that is {\it transmitted through the chain} $c$ that connects the vertex $\tilde{s}$ and the vertex $i$. As a result of this transmission of $v_{\tilde{s}}$, by traversing the chain $c$, we obtain the {\it contribution} $x_{i,c}^{(k+1)}$ into the whole value of $x_i^{(k+1)}$ by this chain. There must be chains through which the non-zero value $v_{\tilde{s}}$ is transmitted to the vertex $i$ to have $x_i^{(k+1)}\!\neq\!0$. 

So, for $\tilde{s}\!\in\!\mathcal{F}^{(k)}$ and $c\!\in C(\tilde{s},i)$ we define $x_{i,c}^{(k+1)}$ as the contribution of the value $v_{\tilde{s}}$ which transmitted along the chain $c\!\in C(\tilde{s},i)$ from $\tilde{s}$ to $i$: 
$$x_{i,c}^{(k+1)}=v_{\tilde{s}}\cdot (-d)^{\ell(c)},\eqno(8)$$
where $$v_{\tilde{s}}=\left\{
\begin{array}{ll}
x_{\tilde{s}}^{(k+1)},& \mbox{ if $c$ is a chain of type (I)};\\
x_{\tilde{s}}^{(k)},& \mbox{ if $c$ is a chain of type (II)}.\\
\end{array}
\right.$$

Traversals along individual chains in (7) computationally performs by the following algorithm. 

\begin{codebox}
\Procname{$\proc{\bf Chain traversal algorithm}\ (c, v_{\tilde{s}}):x_{i,c}^{(k+1)};$}
\li $c'\leftarrow\varnothing$;
\li $x_{i_0,c'}^{(k+1)}\leftarrow v_{\tilde{s}}$;
\li \For $t\!=\!\overline{1,\ell(c)}$:
\Do 
\li $c''\leftarrow c'+(i_{t-1},i_{t})$;
\li $x_{i_t,c''}^{(k+1)}\leftarrow x_{i_{t-1},c'}^{(k+1)}\cdot (-d)$;
\li $c'\leftarrow c''$;
\End
\li \mbox{\bf result}$\ \leftarrow x_{i_t,c'}^{(k+1)}$.
\end{codebox}

We can decompose the set of all computations of an iteration (7) to computations of individual chains' contributions which implemented by $\proc{\bf Chain traversal}$ $\proc{\bf algorithm}$. Using this decomposition, we can implement computations for an iteration of the form (7) in parallel.

For $\tilde{s}\!\in\!\mathcal{F}^{(k)}$, we define $x_{i(\tilde{s})}^{(k+1)}$ as $$x_{i(\tilde{s})}^{(k+1)}=\sum\limits_{c\in C(\tilde{s},i)}x_{i,c}^{(k+1)}.\eqno(9)$$ If $C(\tilde{s},i)\!=\!\varnothing$ then $x_{i(\tilde{s})}^{(k+1)}\!=\!0$ since there are no chains, through which the contributution $v_{\tilde{s}}$ can be transmitted to vertex $i$ from vertex $\tilde{s}$. The value $x_{i(\tilde{s})}^{(k+1)}$ is the sum of all contributions to the value of $x_i^{(k+1)}$ which are transmitted through all chains in $C(\tilde{s},i)$ at the course of the $(k+1)$-th iteration (7). 

\bigskip

Note that during implementation of the iterartions (7), reaching vertices from $\mathcal{F}^{(k+1)}$ can be done not only along simple chains. Traversal of a chain of the type (II) may cause to correct chain that otgoes from some vertex $j$ already reached at the course of an iteration (7), at the same iteration will repeatedly pass through the vertex $\tilde{s}$ from which $j$ was reached. The value obtained along traversals of a such loops will be added to the initial value of $x_{\tilde{s}}^{(k)}$, further transmitted along the chains of $C(\tilde{s},i)$. That is, it will be included in $v_{\tilde{s}}$ from (8) while $\tilde{s}$ is revisited. This happens when $\tilde{s}\!>\!j$.

We show such situation on the Fig.\! 1.1.a) and  Fig.\! 1.1.b). The equations (7) for these graph with starting vertex $2$ are the following: $$\left\{
\begin{array}{l}
x_1^{(k+1)}=x_2^{(k)}(-d),\\
x_2^{(k+1)}=\left (-1+x_1^{(k+1)}+x_3^{(k)}\right )(-d),\\
x_3^{(k+1)}=x_2^{(k+1)}(-d).\\
\end{array}\right.$$

For the 1-st iteration, $x_1^{(0)}\!=\!0$, $x_2^{(0)}\!=\!d$, $x_3^{(0)}\!=\!0$, thus $$x_3^{(1)}=x_2^{(1)}(-d)=\left (-1+x_1^{(1)}\right )d^2=$$ $$=\left (-1+x_2^{(0)}(-d)\right )d^2=(-1-d^2)d^2=-d^2-d^4.$$ The contribution in $x_3^{(1)}$ of $v_s\!=\!d$ transmitted by the cycle $\{2,1,2\}$ gives us $-d^4$ for the whole walk $c_1\!=\!\{2,1,2,3\}$ at the $1$-st iteration. Adding it with the contribution in $x_3^{(1)}$ which is transmitted by the simple chain $c_2\!=\!\{1,3\}$ consisted of one edge $(1,3)$, we obtain the value of $x_3^{(1)}$ at the 1-st iteration as the sum of contributions in $x_3^{(1)}$ transmitted from the vertex $2$ by $3$ the walk $c_1$ and by the simple chain $c_2$: $$x_3^{(1)}=x_{3,c_1}^{(1)}+x_{3,c_2}^{(1)}=-d^4-d^2.$$

Note that, after passing through such a loop (cycle), the contribution equally propagate over all simple chains outgoing from the vertex.

\smallskip

For the graph in Fig. 1.2, such loops (cycles) are absent in the case of the starting vertex $1$. The figure shows how the contributions added to the vertex $5$ along two correct chains that are traversed during one CSS iteration. These are the chains $c_1\!=\!\{1,2,5\}$ and $c_2\!=\!\{1,3,4,5\}$. We have $$x_5^{(1)}=x_{5,c_1}^{(1)}+x_{5,c_2}^{(1)}=d^3-d^4.$$

%

\begin{figure}[htbp]
 \centering
\includegraphics[width=80mm]{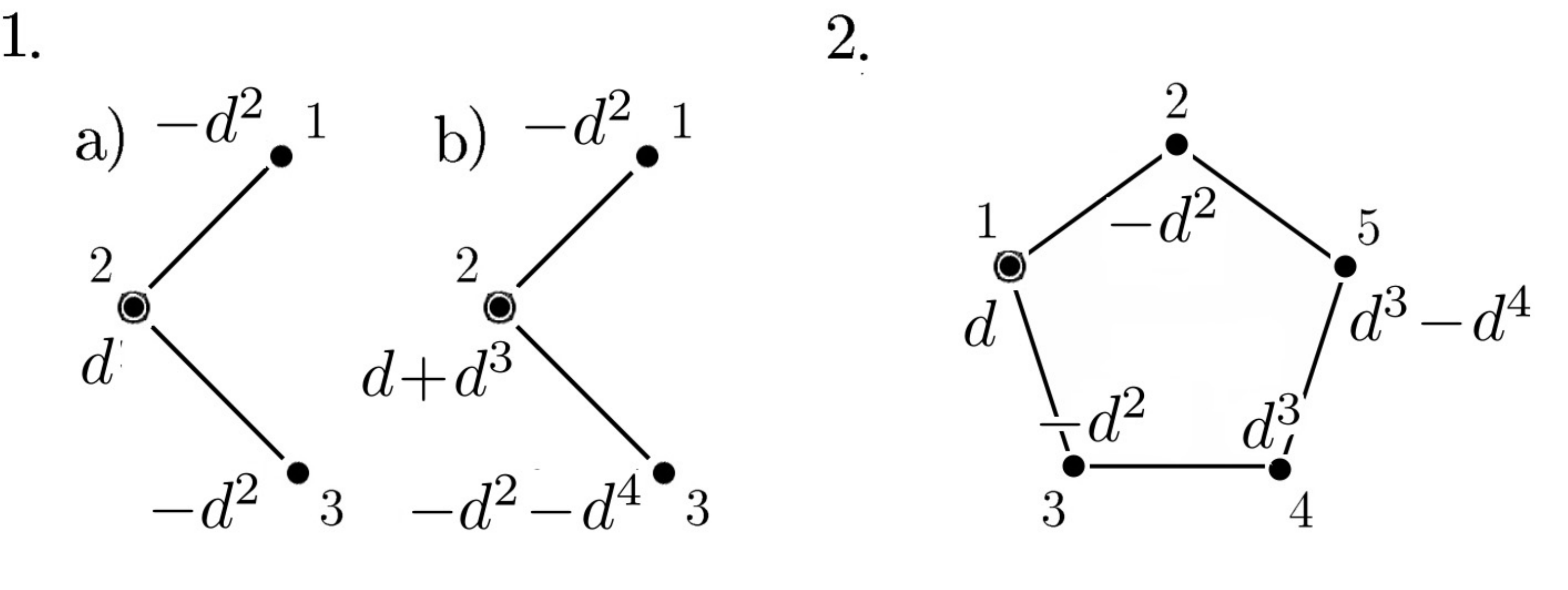}
\caption{Contributions of walks and chains.}
\end{figure}

\subsection{Gauss-Seidel iterations as implementation\\ of the combinatorial CCS}

For $i\!\neq\!s$, we have from (7) that $$x_i^{(k+1)}=\biggl (\sum\limits_{j=1}^{i-1}a_{ij}x_j^{(k+1)}+\sum\limits_{j=i+1}^na_{ij}x_j^{(k)}\biggr )(-d)=$$ $$=\sum\limits_{\substack{\text{$j\!<\!i$}\\ \text{$(i,j)\!\in E$}}}x_j^{(k+1)}\cdot(-d)+\sum\limits_{\substack{\text{$j\!>\!i$}\\ \text{$(i,j)\!\in E$}}}x_j^{(k)}\cdot(-d).$$ That is 
$$x_i^{(k+1)}=\sum\limits_{\substack{\text{$j\!<\!i$}\\ \text{$(i,j)\!\in E$}}}x_j^{(k+1)}\cdot(-d)+\sum\limits_{\substack{\text{$j\!>\!i$}\\ \text{$(i,j)\!\in E$}}}x_j^{(k)}\cdot(-d).\eqno(10)$$

Lemma 2 states that obtaining the vector $x^{(k+1)}$ at the $(k+1)$-th iteration (7) is equivalent to obtaining it using all simple chains from $C(\tilde{s},i)$ in (9) for all $\tilde{s}\!\in\!\mathcal{F}^{(k)}$. 

Proving the following lemma and theorem, we do not consider the situation of having cycles starting and ending at the vertex $\tilde{s}$, an example of which is given above. In this case, the contribution transmitted further after passing through such a loop (cycle) will equally transmitted over all correct simple chains outgoing from the vertex after passing the vertex again. This does not change the equality or inequality to zero of the sum of contributions transmitted along the chains in the course of (7). And therefore it does not change the equality or inequality to zero of $x_i^{(k+1)}$.

\bigskip

\noindent{\bf Lemma 2.} {\it Let $i\!\in\!\mathcal{F}^{(k+1)}$ for the combinatorial CCS, and $x_i^{(k+1)}$ is the value computed at the $(k+1)$-th iteration (7). Then} $$x_{i}^{(k+1)}=\sum\limits_{\tilde{s}\in\mathcal{F}^{(k)}} x_{i(\tilde{s}) }^{(k+1)}=\sum\limits_{\tilde{s}\in\mathcal{F}^{(k)}}\sum\limits_{c\in C(\tilde{s},i)} x_{i,c}^{(k+1)}=\sum\limits_{\tilde{s}\in\mathcal{F}^{(k)}}\sum\limits_{c\in C(\tilde{s},i)}v_{\tilde{s}}\cdot(-d)^{\ell(c)},\eqno(11)$$
{\it where $v_{\tilde{s}}\!=\!x_{\tilde{s}}^{(k)}$ is the value transmitted from the vertex $\tilde{s}\!\in\!\mathcal{F}^{(k)}$ by chains from $C(\tilde{s},i)$.}
\bigskip

\noindent{\bf Proof.} We shall prove Lemma~2 by induction on the number of iterations of the combintorial CCS and iterations of the form (7). We prove the basis of induction and the inductive assumption by induction also. We perform it by induction on length of chains from $C(\tilde{s},i)$ for $\tilde{s}\!\in\!\mathcal{F}^{(k)}$ and $i\!\in \!\mathcal{F}^{(k+1)}$.

\smallskip 

Let us show that Lemma~2 is true for the $1$-st iteration, that is, for $k\!=\!1$. The frontier at the $1$-st iteration consists of a single starting vertex $s$: $\mathcal{F}^{(0)}\!=\!\{s\}$. Let $i\!\in\!\mathcal{F}^{(1)}$ be such that $\ell(s,i)\!=\!1$. Then $C(s,i)$ consists of a single chain $c\!=\!\{s,i\}$ of length $1$, i.e., it consists of the one edge $(s,i)\!\in\!E$. Thus we have $x_{i,c}^{(1)}\!=\!-d^2$ by (8), since $v_s\!= d$ at the $1$-st iteration (7).

For the $1$-st iteration (7), we have $x_j^{(0)}\!=\!0$ for all $j\!>\!i$, $(j,i)\!\in\!E$, $j\!\neq s$. We also have $x_j^{(1)}\!=\!0$ for $j\!<\!i$, $(j,i)\!\in\!E$, $j\!\neq\!s$. Let us assume that $x_j^{(1)}\!\neq\!0$ for such $j$. Consequently, there is a chain of the type (I) or type (II) connected vertices $s$ and $j$ such that, after substitutions of $x_s^{(0)}\!=\!d$ into equations (7) with numbers equal to numbers of vertices of such chain, we obtain $x_j^{(1)}\!\neq\!0$. But, in this case, we have $\ell(s,i)\!>\!1$ since $(j,i)\!\in\!E$ which contradicts assumption that $\ell(s,i)\!=\!1$.

So all the terms in (10) are zero except $x_s^{(0)}$, if $i\!<\!s$, or except $x_s^{(1)}$, if $i\!>\!s$, and we have $$x_i^{(1)}=x_{i(s)}^{(1)}=x_{i,c}^{(1)}=-d^2,$$ where $c\!=\!\{s,i\}$. Thus, it is shown that Lemma~2 is true for vertices $i$ reached at the $1$-st iteration (7) along chains of length $1$.

\smallskip

Suppose that Lemma~2 is true at the $1$-st iteration of (7) for all such $i\!\in\!\mathcal{F}^{(1)}$ that $\ell(s,i)\!=\!l$. Let us show that it will also be true for all such $i\!\in\!\mathcal{F}^{(1)}$ that $\ell(s,i)\!= l+1$.

\smallskip

Since at the $1$-st iteration $x_j^{(0)}\!=\!0$ for $j\!>\!i$, $j\!\neq\!s$, then from (10) we have: $$x_i^{(1)}=\sum\limits_{\substack{\text{$j\!<\!i, j\neq s,$}\\ \text{$(i,j )\!\in E$}}}x_j^{(1)}\cdot (-d)+\alpha\cdot v_s\cdot(-d),\eqno(12)$$ where $\alpha\!=\!1$ if $(s,i)\!\in\!E$, $s\!>\!i$, and in this case we have $\{s,i\}\!\in\!C(s,i)\!\neq\!\varnothing$. If $\alpha\!=\!0$ and there are $x_j^{(0)}\!\neq\!0$, $j\!\neq\!s$, then there are other chains in $C(s,i)$. For all such $j$ we have $C(s,j)\!\neq\!\varnothing$, and, for $c\!\in\!C(s,j)$, $\ell(c)\!\le\!\ell(s,j)\!\le\!l$ since $\ell(s,i)\!=\!l+1$. Thus, by induction asssumption on $l$, we have: $$x_j^{(1)}=\sum\limits_{c'\in C(s,j)}x_{j,c'}^{(1)}.\eqno(13)$$ Since $x_{i, c}^{(1)}\!=\!x_{j, c'}^{(1)}\cdot(-d)$ for $c'\!\in C(s,j)$ such that $c\!=\!c'+(j,i)$, $c\!\in C(s,i)$, then, substituting (13) into (12) and having $v_s\cdot(-d)\!=\!x_{i,c}^{(1)}$ for $c\!=\!\{s,i\}$, we obtain $$x_i^{(1)}=\sum\limits_{c\in C(s,i)}x_{i,c}^{(1)}=x_{i(s)}^{(1)}\eqno(14)$$ for both possible values of $\alpha$.

Thus, for the $1$-st iteration, Lemma~2 is true for vertices $i\!\in\!\mathcal{F}^{(1)}$ reached along chains in $C(\tilde{s},i)$ of any length.

\bigskip 

Assuming that Lemma~2 is true for all iterations up to the $k$-th, let us show that it will also be true at the $(k+1)$-th iteration.

\smallskip

Consider $i\!\in\!\mathcal{F}^{(k+1)}$ such that $\ell(\tilde{s},i)\!=\!1$ for a vertex $\tilde {s}\!\in\!\mathcal{F}^{(k)}$. For this case, $C(\tilde{s},i)$ consists of one chain that is the edge $(\tilde{s},i)$ which, in (10), corresponds to a non-zero term of the form $$x_{i,c}^{(k+1)}=x_{\tilde{s}}^{(k+1)}\cdot(-d)$$ if $\tilde{s}\!<\!i$, or $$x_{i,c}^{(k+1)}=x_{\tilde{s}}^{(k)}\cdot(-d)$$ if $\tilde{s}\!>\!i$. Therefore we have: $$x_i^{(k+1)}=\sum\limits_{\tilde{s}\in\mathcal{F}^{(k)}}\sum\limits_{c\in C(\tilde{s},i)}x_{i,c}^{(k+1)}=\sum\limits_{\tilde{s}\in\mathcal{F}^{(k)}}x_{i(\tilde{s})}^{(k+1)}.$$ That is, Lemma~2 is true for $i\!\in\!\mathcal{F}^ {(k+1)}$ such that $\ell(\tilde{s},i)\!=\!1$ for $\tilde{s}\!\in\!\mathcal{F}^{(k)}$.

\smallskip

Suppose that it is true for all $i\!\in\!\mathcal{F}^{(k+1)}$ such that $\ell(\tilde{s},i)\!\le l$ for $\tilde{s}\!\in\!\mathcal{F}^{(k)}$. Let us show that it will also be true for all $i\!\in\!\mathcal{F}^{(k+1)}$ such that $\ell(\tilde{s},i)\!=\!l+1$. Carrying out induction on $l$ we show that for all vertices reached at the $(k+1)$-th iteration, the statement of the lemma is true.

\smallskip

Let $\mathcal{F}_1^{(k)}\!=\!\{\tilde{s}\in\mathcal{F}^{(k)}\, |\, \tilde{s}<i\}$, $\mathcal{F}_2^{(k)}\!=\!\{\tilde{s}\in\mathcal{F}^{(k)}\, |\, \tilde {s}>i\}$, $\mathcal{F}_1^{(k)}\cup\mathcal{F}_2^{(k)}\!=\!\mathcal{F}^{(k)}$. Consider the first term $S_1$ in the right-hand side of (10): $$S_1=\sum\limits_{\substack{\text{$j\!<\!i,$}\\ \text{$(i,j) \in E$}}}x_j^{(k+1)}\cdot(-d).$$ If $j\!<\!i$ and $x_j^{(k+1)}\!\neq\!0$, then $C(\tilde{s},j)\!\neq\!\varnothing$ since, by the induction assumption on $k$, only components for vertices reached at iterations up to the $k$-th can be nonzero in state vector, and to obtain a nonzero value $x_j^{(k+1)}$ in the right side of (10), correct chains connecting such vertices and $j$ are needed. The length of each chain in $C(\tilde{s},j)$ is less than or equal to $l$, since otherwise $\ell(\tilde{s},i)\!>\!l+1$. 

Therefore, by induction on $l$, we have $$S_1=\sum\limits_{\tilde{s}\in\mathcal{F}_1^{(k)}}\sum\limits_{c\in C(\tilde{s},j)}x_{j, c}^{(k+1)}\cdot(-d).$$ Since by (8) we have $x_{j,c}^{(k+1)}\cdot(-d)=x_{i,c'}^{(k+1)}$, where $c'\!=\!c+(j,i)$, then, taking into account (9), we obtain: $$S_1=\sum\limits_{\tilde{s}\in\mathcal{F}_1^{(k)}}\sum\limits_{c\in C(\tilde{s},i)}x_{i,c}^{(k+1)}=\sum\limits_{\tilde{s}\in\mathcal{F}_1^{(k)}}x_{i(\tilde{s})}^{(k+1)}.\eqno(15)$$ 

Consider the second term $S_2$ in the right-hand side of (10): $$S_2=\sum\limits_{\substack{\text{$j\!>\!i,$}\\ \text{$(i,j) \in E$}}}x_j^{(k)}\cdot(-d).$$ By induction assumption on $k$, for $j\!>\!i$, we have $x_j^{(k)}\!\neq\!0$ if $j\!\in\!\mathcal{F}_2^{(k)}$. Therefore, $$S_2=\sum\limits_{\tilde{s}\in\mathcal{F}_2^{(k)}}x_{i(\tilde{s})}^{(k+1)}\eqno(16)$$ since $x_{\tilde{s}}^{(k)}\cdot(-d)\!=\!x_{i,c}^{(k+1)}\!=\!x_{i(\tilde{s})}^{(k+1)}$ for chains of the form $c\!=\!\{\tilde{s},i\}$, $\ell(c)\!=\!1$, $(\tilde{s},i)\!\in\!E$.

Adding $S_1$ and $S_2$, we obtain from (15) and (16) that $$x_i^{(k+1)}=S_1+S_2=\sum\limits_{\tilde{s}\in\mathcal{F}_1^{(k)}}x_{i(\tilde{s})}^{(k+1)}+\sum\limits_{\tilde{s}\in\mathcal{F}_2^{ (k)}}x_{i(\tilde{s})}^{(k+1)}=\sum\limits_{\tilde{s}\in\mathcal{F}^{(k)}}x_{i(\tilde{s})}^{(k+1)}.$$ This means that Lemma~2 is true at the $(k+1)$-th iteration for all $i\!\in\!\mathcal{F}^{(k+1)}$ such that $\ell(\tilde{s},i)\!=\!l+1$. Thus, by induction on $l$, Lemma~2 is true for all vertices in $\mathcal{F}^{(k+1)}$, i.e., it is true for the $(k+1)$-th iteration. It means that, by the induction on $k$, it is true for an arbitrary $k$ ($k\!\le\!n$). $\qed$

\bigskip

\noindent{\bf Theorem 2.} {\it For a given graph $G$ and starting vertex $s$, the combinatorial CCS and Gauss-Seidel iterations with $b\!=\!e_s$ and $x^{(0)}\!=\!d\cdot e_s$ give the same traversal}.

\bigskip

\noindent{\bf Proof.} To prove Theorem 2, it is necessary to show that, in the course of implementing iterations (7), the condition (5) for the vertex $i\!\in V$ will be met if and only if $i\!\in\!\mathcal{F}^{(k+1)}$ for the combinatorial CCS.

By the proven Lemma~2, $$x_i^{(k+1)}=\sum\limits_{\tilde{s}\in\mathcal{F}^{(k)}}\sum\limits_{c\in C(\tilde{s},i)}v_{\tilde{s}}\cdot(-d)^{\ell(c)},$$ where $v_{\tilde{s}}\!= O(d^{\ell(s,\tilde{s})})\!\neq 0$, and where $\ell(s,\tilde{s})$ is the maximum length of a simple chain connecting the starting vertex $s$ and the vertex $\tilde{s}\!\in\!\mathcal{F}^{(k)}$. By definition (8), when traversing along each edge of a such chain, the value $v_{\tilde{s}}$ which is a sum of powers $d$ taken with some coefficients, is multiplied by $-d$. In this case, the same powers of $d$ in the resulting values $x_{i,c}^{(k+1)}$ have the same sign, since they obtained by the same number of multiplications by $-d$ implemented by $\proc{\bf Chain traversal algorithm}$ for chains of the same lengths. Therefore, if $C(\tilde{s},i)\!\neq\!\varnothing$, then $x_i^{(k+1)}\!\neq\!0$ and condition (5) is met for $i\!\in\!\mathcal{F}^{(k+1)}$ of the combinatorial CCS after the $(k+1)$-th iteration. 

Let $i\!\in V\setminus \mathcal{C}^{(k+1)}$ for the combinatorial CCS, that is, the vertex $i$ is not reached after the $(k+1)$-th iteration. This means that $C(\tilde{s},i)\!=\!\varnothing$ for every $\tilde{s}\!\in\!\mathcal{F}^{(k)}$. By the proven Lemma~2, it follows from this that $$x_i^{(k+1)}=\sum\limits_{\tilde{s}\in\mathcal{F}^{(k)}}x_{i(\tilde {s})}^{(k+1)}=0,$$ since we have $x_{i(\tilde{s})}^{(k+1)}\!=\!0$ if $C(\tilde{s},i)\!=\!\varnothing$. So the condition (5) is not met for $i\!\in\!V\setminus\mathcal{C}^{(k+1)}$.

\smallskip

Thus, each iteration of the combinatorial CCS will produce the same frontiers $\mathcal{F}^{(k)}$, $k\!=\!1, 2, \ldots$, as the iterations (7) produce. This means that we obtain the same graph traversals during execution of the iterations. $\qed$

\section{Examples of graph traversals\\ implemented by~Jacobi and Gauss-Seidel iterations}

For all examples considered in this sections, the starting vertex is vertex $1$. $d\!=\!2$.

\smallskip

Let us compare the traversals implemented by BFS and CCS for the graph $G$ presented at Fig.~2. For this graph, the~transformation $\mbox{\bf F}$ of the state vectors $x^{(k)}$ is given by the following equations of the form (6): $$\left\{
\begin{array}{l}
x_1^{(k+1)}=\left (-1+x_2^{(k)}\right )(-d),\\
x_2^{(k+1)}=\left (x_1^{(k)}+x_3^{(k)}+x_6^{(k)}\right )(-d),\\
x_3^{(k+1)}=\left (x_2^{(k)}+x_4^{(k)}+x_7^{(k)}\right )(-d),\\
x_4^{(k+1)}=\left (x_3^{(k)}\right )(-d),\\
x_5^{(k+1)}=\left (x_6^{(k)}\right )(-d),\\
x_6^{(k+1)}=\left (x_2^{(k)}+x_5^{(k)}+x_7^{(k)}\right )(-d),\\
x_7^{(k+1)}=\left (x_3^{(k)}+x_6^{(k)}+x_8^{(k)}\right )(-d),\\
x_8^{(k+1)}=\left (x_7^{(k)}\right )(-d).\\
\end{array}\right.\eqno(17)$$


\begin{figure}[htbp]
 \centering
\includegraphics[width=30mm]{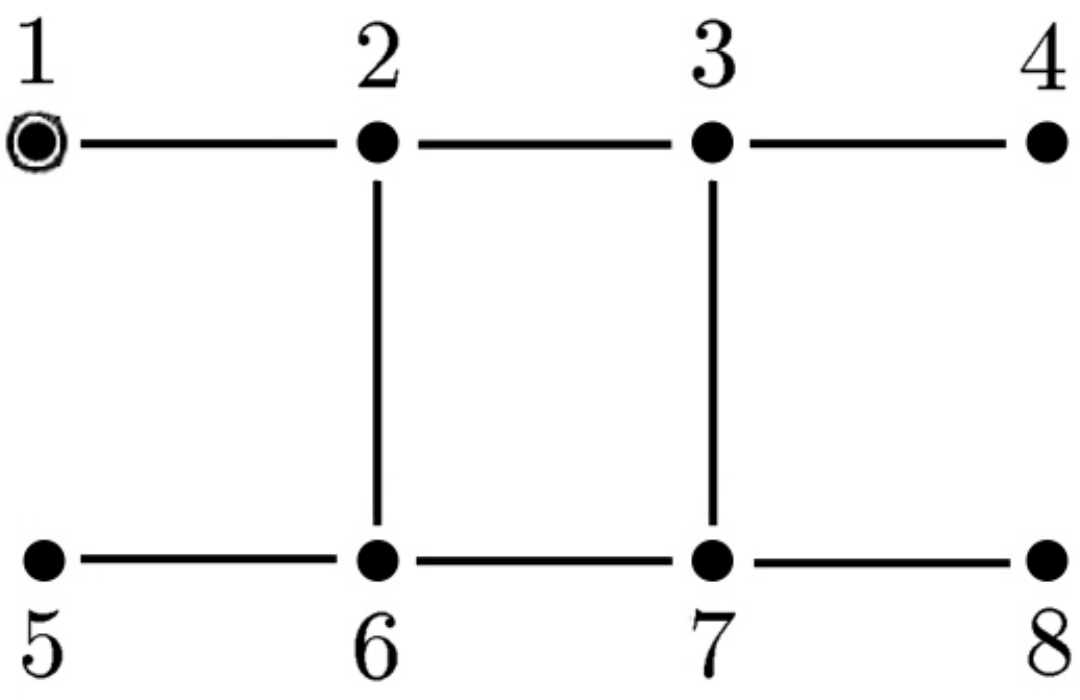}
\caption{Graph on 8 vertices.}
\end{figure}

\noindent Iterations of the traversal, which the iterations gives for the graph, are presented at Fig.~3. It takes 4 iterations to complete the traversal of the graph using BFS, i.e., Jacobi iterations (6).

For an CCS iteration, the transformation $\mbox{\bf F}$ of the state vectors $x^{(k)}$ is given by the equations (18): $$\left\{
\begin{array}{l}
x_1^{(k+1)}=\left (-1+x_2^{(k)}\right )(-d),\\
x_2^{(k+1)}=\left (x_1^{(k+1)}+x_3^{(k)}+x_6^{(k)}\right )(-d),\\
x_3^{(k+1)}=\left (x_2^{(k+1)}+x_4^{(k)}+x_7^{(k)}\right )(-d),\\
x_4^{(k+1)}=\left (x_3^{(k+1)}\right )(-d),\\
x_5^{(k+1)}=\left (x_6^{(k)}\right )(-d),\\
x_6^{(k+1)}=\left (x_2^{(k+1)}+x_5^{(k+1)}+x_7^{(k)}\right )(-d),\\
x_7^{(k+1)}=\left (x_3^{(k+1)}+x_6^{(k+1)}+x_8^{(k)}\right )(-d),\\
x_8^{(k+1)}=\left (x_7^{(k+1)}\right )(-d).\\
\end{array}\right.\eqno(18)$$


\begin{figure}[htbp]
 \centering
\includegraphics[width=80mm]{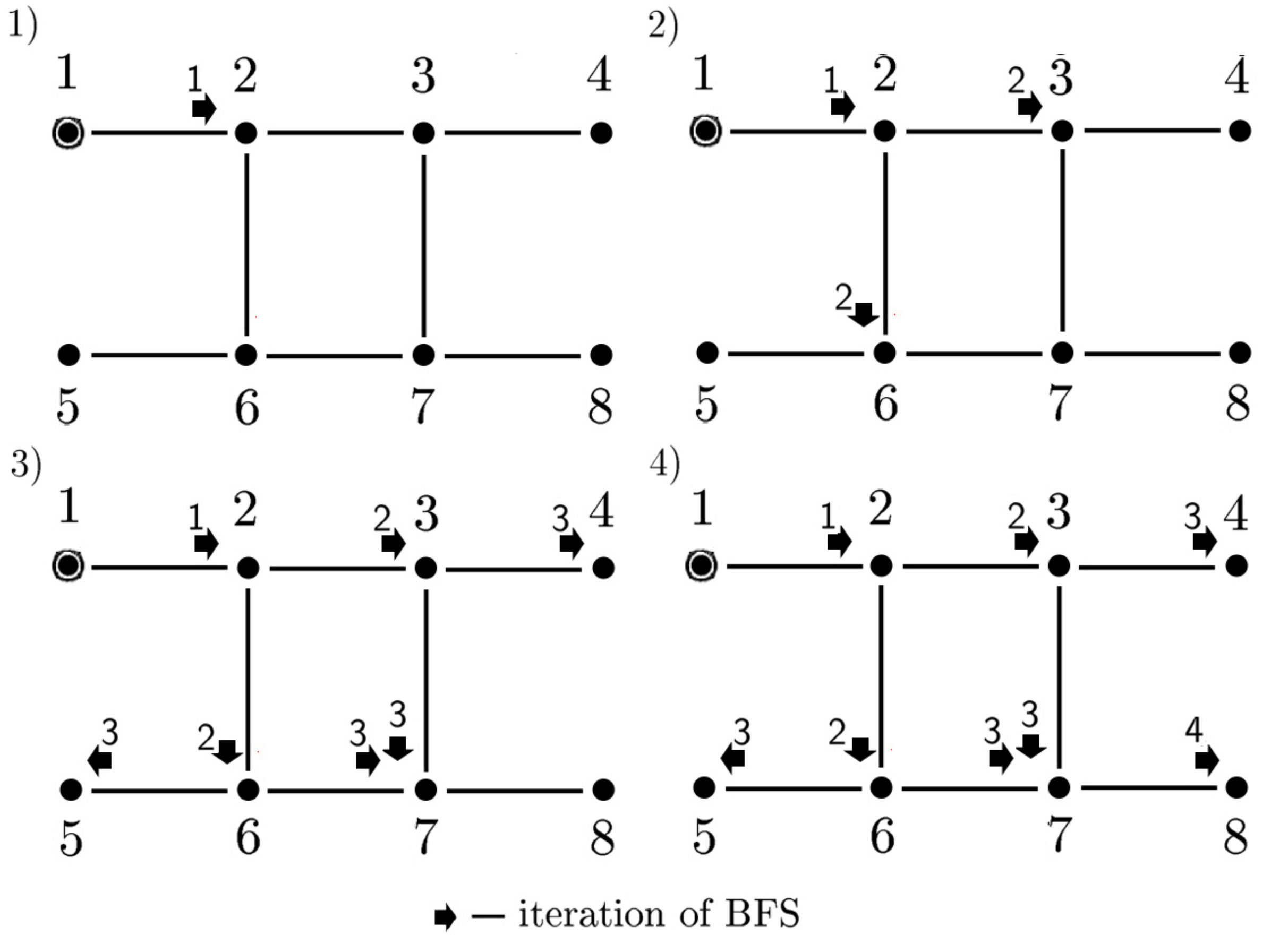}
\caption{Graph traversal implemented by BFS.}
\end{figure}


\begin{figure}[htbp]
 \centering
\includegraphics[width=80mm]{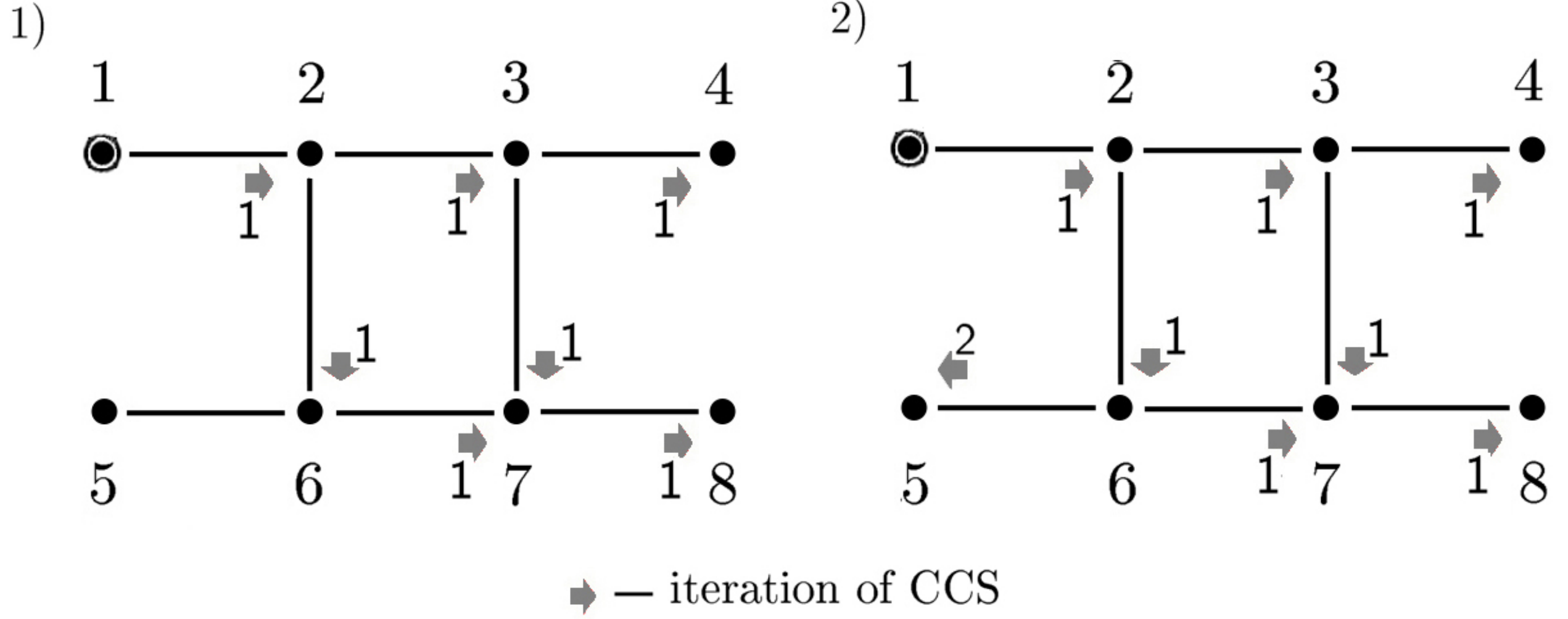}
\caption{Graph traversal implemented by CCS.}
\end{figure}

Implementing iterations (18), we obtain a traversal of the graph just for 2 iterations. As we may see, unlike BFS, when performing the~first iteration (18) of CCS, a transition will be made along correct chains outgoing from vertex $2$ which is reached from vertex $1$ at the same iteration (Fig.~4).

The~state vectors $x^{(k)}$ and the sets $\mathcal{C}^{(k)}$ of vertices reached after the $k$-th iterations of BFS and CCS implemented by iterations (17) and (18) are listed below.  

\noindent\underline{\it BFS (implemented by (17))}:

\smallskip
 
\noindent Initialization: $x^{(0)}=(2, 0, 0, 0, 0, 0, 0, 0)$, $\mathcal{F}^{(0)}\!=\!\{1\}$, $\mathcal{C}^{(0)}\!=\!\{1\}$.\\
\noindent 1) $x^{(1)}\!=\!(2, -4, 0, 0, 0, 0, 0, 0)$, $\mathcal{F}^{(1)}\!=\!\{2\}$, $\mathcal{C}^{(1)}\!=\!\{1, 2\}$;\\
\noindent 2) $x^{(2)}\!=\!(10, -4, 8, 0, 0, 8, 0, 0)$, $\mathcal{F}^{(2)}\!=\!\{3,6\}$, $\mathcal{C}^{(2)}\!=\!\{1, 2, 3, 6\}$;\\
\noindent 3) $x^{(3)}\!=\!(10, -52, 8, -16, -16, 8, -32, 0)$, $\mathcal{F}^{(3)}\!=\!\{5,7\}$, $\mathcal{C}^{(3)}\!=\!\{1, 2, 3, 4, 5, 6, 7\}$;\\
\noindent 4) $x^{(4)}\!=\!(106, -52, 200, -16, -16, 200, -32, 64)$, $\mathcal{F}^{(4)}\!=\!\{8\}$, $\mathcal{C}^{(4)}\!=\!\{1, 2, 3, 4, 5, 6, 7, 8\}$.

\smallskip

\noindent\underline{\it CCS (implemented by (18))}:

\smallskip
 
\noindent Initialization: $x^{(0)}\!=\!(2, 0, 0, 0, 0, 0, 0, 0)$, $\mathcal{F}^{(0)}\!=\!\{1\}$, $\mathcal{C}^{(0)}\!=\!\{1\}$.\\
\noindent 1) $x^{(1)}\!=\!(2, -4, 8, -16, 0, 8, -32, 64)$, $\mathcal{F}^{(1)}\!=\!\{2, 3, 4, 6, 7, 8\}$, $\mathcal{C}^{(1)}\!=\!\{1, 2, 3, 4, 6, 7, 8\}$;\\
\noindent 2) $x^{(2)}\!=\!(10, -52, 200, -400, -16, 200, -928, 1856)$, $\mathcal{F}^{(2)}\!=\!\{5\}$, $\mathcal{C}^{(2)}\!=\!\{1, 2, 3, 4, 5, 6, 7, 8\}$.

\smallskip

For the graph shown at Fig. 5, the transformation $\mbox{\bf F}$ of the state vector $x^{(k)}$ of the form (6) is given by the equations (19): $$\left\{
\begin{array}{l}
x_1^{(k+1)}=\left (-1+x_2^{(k)}\right )(-d),\\
x_2^{(k+1)}=\left (x_1^{(k)}+x_3^{(k)}\right )(-d),\\
x_3^{(k+1)}=\left (x_2^{(k)}+x_4^{(k)}\right )(-d),\\
x_4^{(k+1)}=\left (x_3^{(k)}+x_5^{(k)}\right )(-d),\\
x_5^{(k+1)}=\left (x_4^{(k)}\right )(-d).\\
\end{array}\right.\eqno(19)$$ The transformation $\mbox{\bf F}$ implemented by CCS for the graph is as follows: $$\left\{
\begin{array}{l}
x_1^{(k+1)}=\left (-1+x_2^{(k)}\right )(-d),\\
x_2^{(k+1)}=\left (x_1^{(k+1)}+x_3^{(k)}\right )(-d),\\
x_3^{(k+1)}=\left (x_2^{(k+1)}+x_4^{(k)}\right )(-d),\\
x_4^{(k+1)}=\left (x_3^{(k+1)}+x_5^{(k)}\right )(-d),\\
x_5^{(k+1)}=\left (x_4^{(k+1)}\right )(-d).\\
\end{array}\right.\eqno(20)$$


\begin{figure}[htbp]
 \centering
\includegraphics[width=40mm]{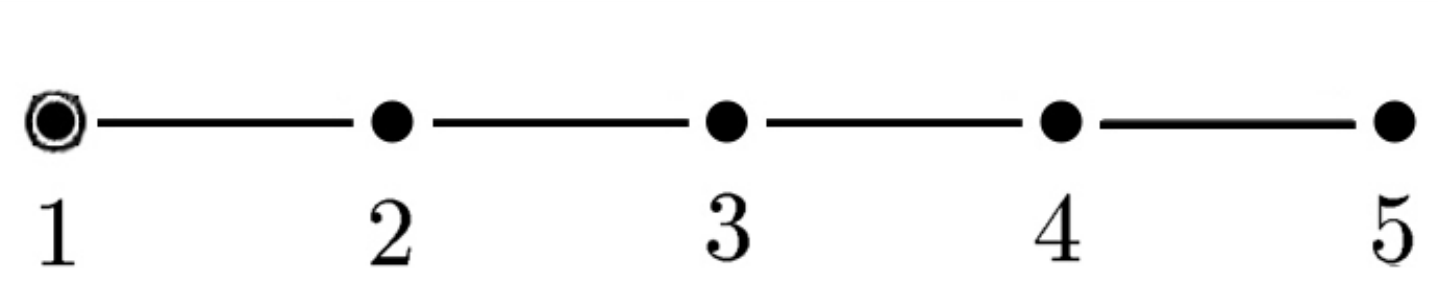}
\caption{Correct chain.}
\end{figure}

For the graph at Fig. 5, we have the correct chain outgoing from vertex $2$, to which all other vertices of this graph belong except the starting one. So all vertices of this graph will be visited at the course of the first iteration of CCS (Fig.~6). 


\begin{figure}[htbp]
 \centering
\includegraphics[width=45mm]{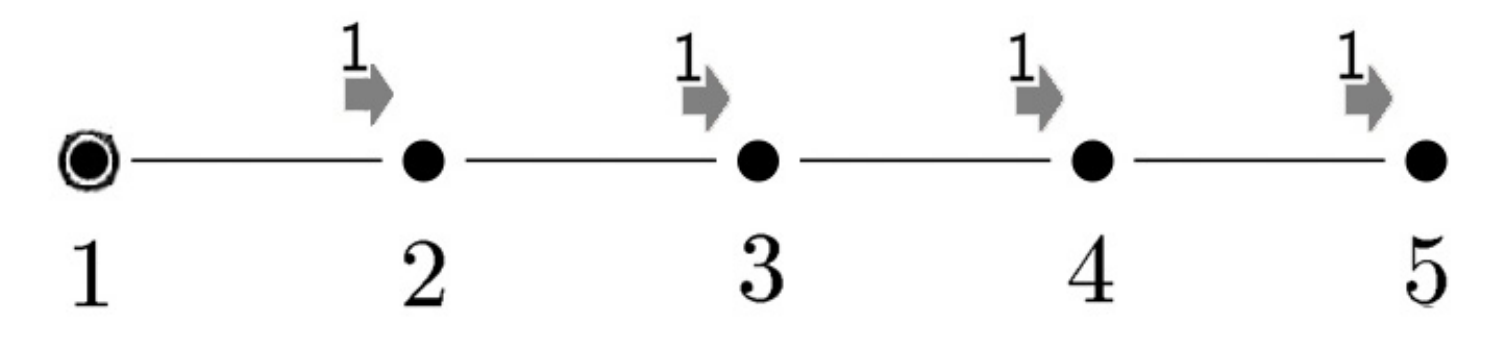}
\caption{Graph traversal implemented by CCS.}
\end{figure}

For~BFS, it takes four iterations which is the maximum possible number of iterations considering all possible starting vertices for a traversal.

\smallskip

\noindent \underline{\it BFS (implemented by (19))}:

\smallskip

\noindent Initialization: $x^{(0)}\!=\!(2, 0, 0, 0, 0)$, $\mathcal{F}^{(0)}\!=\!\{1\}$, $\mathcal{C}^{(0)}\!=\!\{1\}$.\\
\noindent 1) $x^{(1)}\!=\!(2, -4, 0, 0, 0)$, $\mathcal{F}^{(1)}\!=\!\{2\}$, $\mathcal{C}^{(1)}\!=\!\{1, 2\}$;\\
\noindent 2) $x^{(2)}\!=\!(10, -4, 8, 0, 0)$, $\mathcal{F}^{(2)}\!=\!\{3\}$, $\mathcal{C}^{(2)}\!=\!\{1, 2, 3\}$;\\
\noindent 3) $x^{(3)}\!=\!(10, -36, 8, -16, 0)$, $\mathcal{F}^{(3)}\!=\!\{4\}$, $\mathcal{C}^{(3)}\!=\!\{1, 2, 3, 4\}$;\\
\noindent 4) $x^{(4)}\!=\!(74, -36, 104, -16, 32)$, $\mathcal{F}^{(3)}\!=\!\{5\}$, $\mathcal{C}^{(4)}\!=\!\{1, 2, 3, 4, 5\}$.

\smallskip

\noindent \underline{\it CCS (implemented by (20))}:

\smallskip

\noindent Initialization: $x^{(0)}\!=\!(2, 0, 0, 0, 0)$, $\mathcal{F}^{(0)}\!=\!\{1\}$, $\mathcal{C}^{(0)}\!=\!\{1\}$.\\
\noindent 1) $x^{(1)}\!=\!(2, -4, 8, -16, 32)$, $\mathcal{F}^{(1)}\!=\!\{2, 3, 4, 5\}$, $\mathcal{C}^{(1)}\!=\!\{1, 2, 3, 4, 5\}$.

\smallskip


\begin{figure}[htbp]
 \centering
\includegraphics[width=40mm]{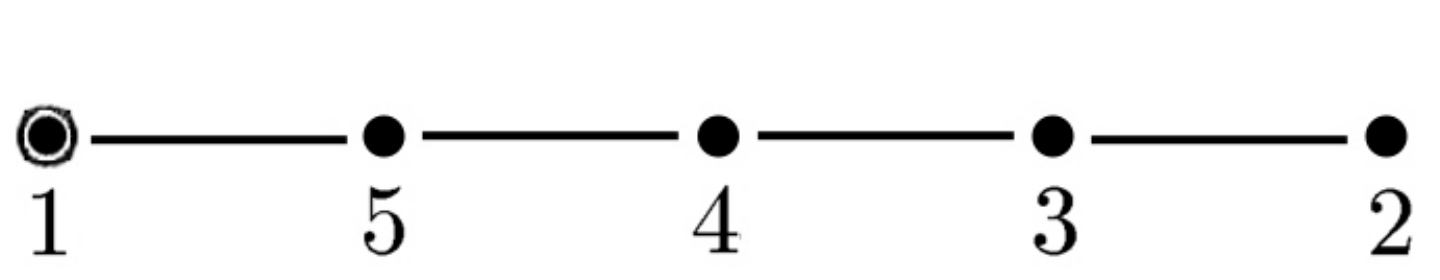}
\caption{Incorrect chain.}
\end{figure}


\begin{figure}[htbp]
 \centering
\includegraphics[width=40mm]{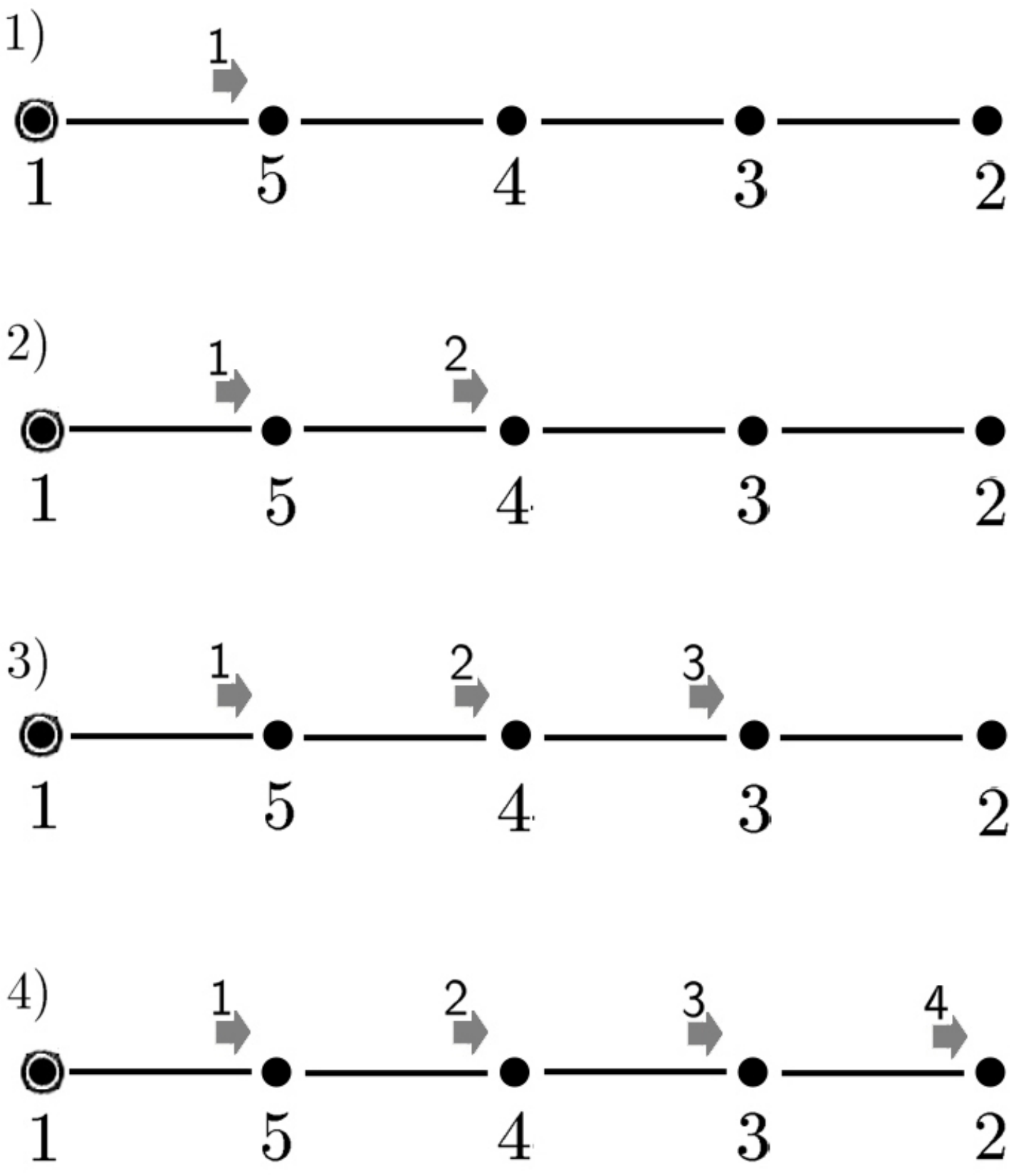}
\caption{Graph traversal implemented by CCS.}
\end{figure}

With different vertex numbering (Fig.~7), at every CCS iteration, there are no correct chains outgoing from visited vertices. Therefore, to reach all vertices of the graph, both BFS and CSS need to carry out 4 iterations. For this case, the traversal implemented by CCS completely repeat the traversal implemented by BFS (Fig.~8).

\section{Unsigned CCS}

The following transformation may be used as the transformation $\mbox{\bf F}$ of the state vector in the algorithm $\proc{\bf Traversal of a connected component}$:
$$x_i^{(k+1)}=\biggl (b_i+\sum\limits_{j=1}^{i-1}a_{ij}x_j^{(k+1)}+\sum\limits_{j=i+1}^{n}a_{ij}x_j^{(k)}\biggr ) d.\eqno(21)$$ Let us call as {\it unsigned CCS} the algorithm $\proc{\bf Traversal of a connected com-}$ $\proc{\bf ponent}$ that uses (21) for $\mbox{\bf F}$.

Unsigned CCS, implementing the same graph traversal as CCS, performs computations using only non-negative integer values if $d\!\in\!\mathbb{Z}$. The absence of division and possible absence of even multiplication for $d\!=\!1$ allows to sagnificantly reduce the time required to implement a graph traversal. 

The transformation (21) is a modification of the transformation $\mbox{\bf F}$ used by CSS. From a transformation of the form $$x^{(k+1)}=(b-Lx^{(k+1)}-U {x}^{(k)})\, d,$$ we move on to a transformation of the form $$x^{(k+1)}=(b+Lx^{(k+1)}+Ux^{(k)})\, d.$$ 

If we do not take into account the multiplication by $d$, the unsigned CCS iteration is similar to the BFS iteration, in which each component of the state vector is computed using components with smaller indices already computed at this iteration. But at the same time, the multiplication (division) by $d$ and getting inverse element for addition may be fundamental to apply the graph traversals to solve some discrete analysis problems. For example, implementation of CCS for perturbed graph matrices gives an approach to solve the graph isomorphism problem~\cite{Prolubnikov1}; it also may be leveraged to perform connectivity test using perturbations of a graph adjacency matrix elements \cite{Prolubnikov2} and to computationally solve other problems on graphs.

\section{State vector regularization and vertex masking} 

Using multiplication instead of division in transformation (7) allows us to reduce the time it takes to implement graph traversal. For stability of computations, to prevent overflow or vanishing to zero of the state vector components as machine numbers if $d\!\in\!(0,1)$, after a given number of iterations $M$, the state vector can be regularized: $$x_i^{(k)}\leftarrow \frac{1}{d^M}\,x_i^{(k)}.\eqno(22)$$

In addition, without changing the order of computational complexity of CCS, a significant acceleration of graph traversal implementation can be achieved using vertex masking. {\it Masking a vertex} means excluding it from further computations. Masked vertices are the vertices from which previously unvisited vertices cannot be reached in the course of next iterations of the traversal algorithm. We can both mask the vertices that belong to already founded connected components of a graph and mask already visited vertices of the currently traversed connected component at an iteration of the $\proc{\bf Finding all connected components}$ algorithm. In the second case, when performing computations at~the $(k+1)$-th iteration~(7), vertices that belonged to frontiers obtained at the previous $k\!-\!1$ iterations of the algorithm $\proc{\bf Traversal of a connected component}$ are masked.

For economy of computer memory and to make computions faster, the computations must be carried out using a {\it portrait} of adjacency matrix. The portrait contains only non-zero elements of the matrix, i.e., it is the list of graph's edges. Using the portrait, implementation of one iteration of the form (6) or (7) has computational complexity of $O(m)$ since we can perform one iteration at one pass through the list of a graph edges.

\section{Traversal of a graph and numbering of its vertices}

The presence of correct chains in a graph is determined by the numbering of the graph vertices. So this  numbering and the choice of the starting vertex are both determine the computational complexity of a graph traversal by CCS. The numbering of a graph's vertices is a mandatory parameter of an instance of the problem on graphs. 

Consider the graph for which it is assumed that information is regularly updated with addition of vertices and edges to the graph or with its deletion from it. Optimization of a numbering of graph vertices can be done by numbering them based on their distances from chosen vertex that is starting vertex. With optimal numbering of vertices, for which there are correct chains that connects starting vertex with any other vertex, to perform a traversal, CCS will have the minimum possible computational complexity for traversal algorithms that is equal to $O(m)$ since we have to do the only one iteration to traverse a graph. To achive this efficiency, we can renumber graph vertices using rechability tree that we obtain after previous executions of~a~traversal algorithm for the graph of a problem which is modified. If the enumeration of the graph vertices is close to the optimal in sense of having correct chains, the computational complexity of CCS also will be of $O(m)$ for the graph.

Note that it may be natural to assign graph vertices numbers which are reflects distances between objects they are represent. This happens, for example, when numbering objects for transport networks. Randomness in such input data as numbering of graph's vertices is inevitable in applications but it can be minimized. A graph with solely random numbering of vertices can be interpreted as a graph of a transport network stored in a database table with randomly generated numbers of records, which may happen in practice only hypothetically with long-term input of data as the network develops. And even for solely random enumerations, there are a lot of instances with correct chains.

\section{The number of iterations required by BFS and CCS\\ to traverse a connected graph}

Let us compare the number of iterations required by BFS and CCS to~traverse a graph. If we do not consider such details of realizations of the traversals as parallelization of~computations and masking of the vertices, the~computational complexity of a graph traversal algorithm for connected graph is determined by the~number of iterations required to visit all its  vertices. 

Let $N_{\mbox{\tiny BFS}}$ be the number of iterations which BFS required to~complete traversal of a connected graph, and let $N_{\mbox{\tiny CCS}}$ be the number of iterations which CCS required for the same graph and the same starting vertex. The values $N_{\mbox{\tiny BFS}}$ and $N_{\mbox{\tiny CCS}}$ are determined by the choice of the starting vertex, but $N_{\mbox{\tiny CCS}}$ is also determined by the numbering of~vertices~in~the~given graph.

Since both BFS and CCS perform the transitions through edges incident the frontiers' vertices, we can assert that for any graph $N_{\mbox{\tiny CCS}}\!\le\!N_{\mbox{\tiny BFS}}$. At the same time, if, in~the course of CCS iterations, we visit at least one vertex, from which at least one correct chain outgoes, then reachability of vertices from the starting one will be determined in fewer iterations than the algebraic BFS would require for it, i.e., $N_{\mbox{\tiny CCS}}\!<\!N_{\mbox{\tiny BFS}}$ in this case. So the~following statement is true.

\bigskip

\noindent {\bf Statement.} {\it For the same starting vertex, CCS will complete its work in a number of iterations that is equal to or less than required for~BFS}.

\section{Computational experiment}

The goal of our experiment was to evaluate the effect of the graph diameter on the difference in the average number of CCS and BFS iterations. To do this, we compute the total numbers of iterations required by CCS and BFS to traverse all graphs from a given set of generated graphs starting from the same starting vertices and find the ratio of these numbers. This ratio is equal to the ratio of the average number of CCS iterations to the average number of BFS iterations needed to traverse a graph. The random graphs used in the experiment are those in which, when they were generated,
\begin{itemize} 
\item[1)] the numbers (labels) of vertices are randomly assigned,
\item[2)] randomly selected non-adjacent vertices are connected by an edge.
\end{itemize}

We define an {\it extended star} as a graph in which one vertex is adjacent to end-vertices of several simple chains. We call these chains {\it rays} of the extended star. To estimate the effect of the diameter on the difference in the average number of CCS and BFS iterations, all graphs used for the experiment were generated based on two types of graphs:
\begin{itemize}
\item[(I)] extended stars with rays of different (random) lengths,
\item[(II)] extended stars with rays of equal length.
\end{itemize}

Edges connecting previously non-adjacent edges were added to the generated extended star graph with random vertex numbering. Let $E(G)$ denote the set of edges of a graph $G$. The algorithm for generating a random graph is as follows.
\smallskip

\noindent 1. We generate on the set of vertices $V\!=\!\{1,\ldots,n\}$ an extended star $G^{(0)}$ of type (I) or type (II).

\smallskip

\noindent 2. Add random edges to $G^{(0)}$: $$E(G^{(i+1)})\rightarrow E(G^{(i)})\cup \{e_{i+1}\},$$ where $e_{i+1}\!\not\in\!E(G^{(i)})$, $i\!=\!\overline{1,\tilde{m}}$, $\tilde{m}$ is the number of random edges added to $G^{(0)}$. The number of edges in the generated graph $G\!=\!G^{(\tilde{m})}$ is $$m\!=\!|E(G^{(\tilde{m})})|\!=\!|E(G^{(0)})|+\tilde{m}\!=\!n-1+\tilde{m}.$$

\smallskip

So the parameters of the graphs used in the experiment are as follows: $n$ is the number of vertices of the extended star, on the basis of which the random graph is generated, $\tilde{m}$ is the number of random edges added to it, characterizing the density of the graph. For graphs generated on the basis of an extended star with rays of the same length, the parameters are as follows: $\ell$ is the length of a ray of the extended star, $r$ is the number of rays.

Note that the denser the graph, i.e. the larger the value of $\tilde{m}$, the smaller, as a rule, the diameter of the generated random graph. Moreover, when adding random edges, the diameter of the resulting graph can decrease very quickly, which can be estimated by the sharp decrease in the number of iterations of both CCS and BFS when adding random edges to the extended star, as shown in the tables below for $\tilde{m}\!=\!0$ and $\tilde{m}\!=\!2n$. In addition, the diameter of a random graph of the type (II) we generate, as a rule, is smaller, the shorter the length $\ell$ of the  ray of the extended star from which the graph is obtained.

\smallskip

Let $N_{\mbox{\tiny CCS}}^{(i)}$ and $N_{\mbox{\tiny BFS}}^{(i)}$ be the numbers of iterations that need to be performed, respectively, using CCS and BFS to traverse the $i$-th graph from a set of $M$ random graphs with given parameters. Let $$\overline{N}_{\mbox{\tiny CCS}}=\sum\limits_{i=1}^MN_{\mbox{\tiny CCS}}^{(i)},$$ $$\overline{N}_{\mbox{\tiny BFS}}=\sum\limits_{i=1}^MN_{\mbox{\tiny BFS}}^{(i)}.$$ Tables 1-4 show the results of experiments. To calculate each ratio $\overline{N}_{\mbox{\tiny CCS}}/\overline{N}_{\mbox{\tiny BFS}}$, $M\!=\!10\,000$ randomly generated graphs of the specified types (I) and (II) with $n\!=\!101$ and other specified parameters were traversed using CCS and BFS.

\begin{table}[h!]
\caption{Random sparse graphs of type (I). $n=101$.}
\begin{center}
\begin{tabular}{|c|c|}
\hline
$\tilde{m}$ & $\overline{N}_{\mbox{\tiny CCS}}/\overline{N}_{\mbox{\tiny BFS}}$\\
\hline
$0$ & $315\,990/619\,604\approx 0.51$\\
\hline
$2n$ & $44\,646/60\,148\approx 0.56$\\
\hline
$5n$ & $30\,288/52\,269\approx 0.58$\\
\hline
$10n$ & $25\,285/40\,431\approx 0.63$\\
\hline
\end{tabular}
\end{center}
\end{table}

\begin{table}[h!]
\caption{Random sparse graphs of type (II). $n=101$, $\ell=50$, $r=2$.}
\begin{center}
\begin{tabular}{|c|c|}
\hline
$\tilde{m}$ & $\overline{N}_{\mbox{\tiny CCS}}/\overline{N}_{\mbox{\tiny BFS}}$\\
\hline
$0$ & $381\,702/752\,844\approx 0.51$\\
\hline
$2n$ & $24\,563/41\,462\approx 0.59$\\
\hline
$5n$ & $20\,007/30\,059\approx 0.67$\\
\hline
$10n$ & $19\,925/23\,346\approx 0.85$\\
\hline
\end{tabular}
\end{center}
\end{table}

\begin{table}[h!]
\caption{Random sparse graphs of type (II). $n=101$, $\ell=20$, $r=5$.}
\begin{center}
\begin{tabular}{|c|c|}
\hline
$\tilde{m}$ & $\overline{N}_{\mbox{\tiny CCS}}/\overline{N}_{\mbox{\tiny BFS}}$\\
\hline
$0$ & $95\,746/154\,515\approx 0.62$\\
\hline
$2n$ & $24\,705/41\,494\approx 0.59$\\
\hline
$5n$ & $20\,005/30\,062\approx 0.67$\\
\hline
$10n$ & $19\,933/23\,360\approx 0.85$\\
\hline
\end{tabular}
\end{center}
\end{table}

\begin{table}[h!]
\caption{Random sparse graphs of type (II). $n=101$, $\ell=10$, $r=10$.}
\begin{center}
\begin{tabular}{|c|c|}
\hline
$\tilde{m}$ & $\overline{N}_{\mbox{\tiny CCS}}/\overline{N}_{\mbox{\tiny BFS}}$\\
\hline
$0$ & $170\,693/304\,627\approx 0.56$\\
\hline
$2n$ & $24\,637/41\,540\approx 0.59$\\
\hline
$5n$ & $20\,005/30\,052\approx 0.67$\\
\hline
$10n$ & $19\,939/23\,429\approx 0.85$\\
\hline
\end{tabular}
\end{center}
\end{table}

For dense graphs with the number of edges equal to $n+n^2/4$, $n\!=\!101$, which is $60\%$ of the number of edges in the complete graph, we have $\overline{N}_{\mbox{\tiny CCS}}/\overline{N}_{\mbox{\tiny BFS}}\!=\!16\,851/20\,000\!\approx \!0.84$. The averaging is obtained over $M\!=\!10\,000$ random graphs.

To calculate ratios $\overline{N}_{\mbox{\tiny CCS}}/\overline{N}_{\mbox{\tiny BFS}}$ presented in Tables~5--9, $M\!=\!1\,000$ randomly generated graphs of the types (I) and (II) with specified parameters were traversed using CCS and BFS.

\begin{table}[h!]
\caption{Random sparse graphs of type (I). $n=1001$.}
\begin{center}
\begin{tabular}{|c|c|}
\hline
$\tilde{m}$ & $\overline{N}_{\mbox{\tiny CCS}}/\overline{N}_{\mbox{\tiny BFS}}$\\
\hline
$0$ & $310\,023/618\,989\approx 0.5$\\
\hline
$2n$ & $3\,140/6\,006\approx 0.52$\\
\hline
$5n$ & $2\,102/4\,046\approx 0.52$\\
\hline
$10n$ & $2\,000/3\,189\approx 0.63$\\
\hline
\end{tabular}
\end{center}
\end{table}

\begin{table}[h!]
\caption{Random sparse graphs of type (II). $n=1001$, $\ell=500$, $r=2$.}
\begin{center}
\begin{tabular}{|c|c|}
\hline
$\tilde{m}$ & $\overline{N}_{\mbox{\tiny CCS}}/\overline{N}_{\mbox{\tiny BFS}}$\\
\hline
$0$ & $376\,913/752\,406\approx 0.5$\\
\hline
$2n$ & $3\,119/6\,001\approx 0.52$\\
\hline
$5n$ & $2\,076/4\,045\approx 0.51$\\
\hline
$10n$ & $2\,000/3\,185\approx 0.63$\\
\hline
\end{tabular}
\end{center}
\end{table}

\begin{table}[h!]
\caption{Random sparse graphs of type (II). $n=1001$, $\ell=200$, $r=5$.}
\begin{center}
\begin{tabular}{|c|c|}
\hline
$\tilde{m}$ & $\overline{N}_{\mbox{\tiny CCS}}/\overline{N}_{\mbox{\tiny BFS}}$\\
\hline
$0$ & $154\,000/298\,341\approx 0.52$\\
\hline
$2n$ & $3\,114/6\,011\approx 0.52$\\
\hline
$5n$ & $2\,079/4\,041\approx 0.51$\\
\hline
$10n$ & $2\,000/3\,220\approx 0.62$\\
\hline
\end{tabular}
\end{center}
\end{table}

\begin{table}[h!]
\caption{Random sparse graphs of type (II). $n=1001$, $\ell=100$, $r=10$.}
\begin{center}
\begin{tabular}{|c|c|}
\hline
$\tilde{m}$ & $\overline{N}_{\mbox{\tiny CCS}}/\overline{N}_{\mbox{\tiny BFS}}$\\
\hline
$0$ & $79\,238/148\,825\approx 0.53$\\
\hline
$2n$ & $3\,122/5\,993\approx 0.52$\\
\hline
$5n$ & $2\,089/4\,063\approx 0.51$\\
\hline
$10n$ & $2\,000/3\,202\approx 0.62$\\
\hline
\end{tabular}
\end{center}
\end{table}

The results of the experiment show that the larger the graph diameter, the greater the average difference in the number of CCS and BFS iterations. On average, for generated sparse graphs with $n\!=\!101$, the difference is from $15$ to $49$ percent of number of BFS iterations needed to traverse a graph (Tables~1--4), and, on average, it is from $38$ to $50$ percent with $n\!=\!1001$ (Tables~5--8). For~dense graphs with $n\!=\!101$, the difference is $16$ percent on average.

\section*{Conclusions}

We consider the methods of simple iterations for solving systems of linear algebraic equations associated with graphs as realizations of graph traversals. These systems are systems with modified adjacency matrices of graphs and with the specified right-hand side. There are two different algorithms for graph traversal can be produced by~this approach. One of them implemented by iterations of the Jacobi method and the other by iterations of the Gauss-Seidel method. The traversal associated with the Jacobi method is equivalent to the breadth first search, and the traversal associated with the  Gauss-Seidel method is~not~equivalent to either the depth-first search or the breadth-first search. For any instance of the problem of finding connected components of a graph, the number of iterations required for the presented graph traversal does not exceed the number of iterations required for the breadth-first search for~the~same given starting vertex. For~a~large number of instances of the problem, fewer iterations will be required for the traversal associated with Gauss-Seidel iterations.

\end{document}